# Dynamics of the securities market in the information asymmetry context: Developing a methodology for emerging securities markets

**Kostyantyn Anatolievich Malyshenko**
Ph. D., Associate Professor, Department "Economy and Finance", V.I. Vernadsky Crimean Federal University, Simferopol, Russian Federation
*Email: docofecon@mail.ru*

**Majid Mohammad Shafiee\***
Ph. D., Associate Professor - Department of Management, University of Isfahan, Isfahan, Iran
*Postal Code: 81746-73441, Phone: +98-0313-793-5227*
*Email: m.shafiee@ase.ui.ac.ir*
\*Corresponding Author

**Vadim Anatolievich Malyshenko**
Ph. D., Associate Professor, Department "Economy and Finance", V.I. Vernadsky Crimean Federal University, Simferopol, Russian Federation
*Email: malyshenko1973@inbox.ru*

**Marina Viktorovna Anashkina**
M.Sc., Department "Economy and Finance", V.I. Vernadsky Crimean Federal University, Simferopol, Russian Federation
*Email: iriska_3640@mail.ru*

**Authors Biography:**
*Malyshenko, Kostyantyn Anatolyevich*, Ph.D. is an Associate Professor in Economics of the enterprise and organization of production. His scientific interest is financial investments, the securities market. He is an associate professor of the Department of Economics and Finance of the Institute of Economics and Management of the Humanitarian Pedagogical Academy (branch) of the Federal State Unitary Enterprise «Crimean Federal University named after V.I. Vernadsky». He is the author of 150 publications, including articles indexed in SCOPUS and WoS.

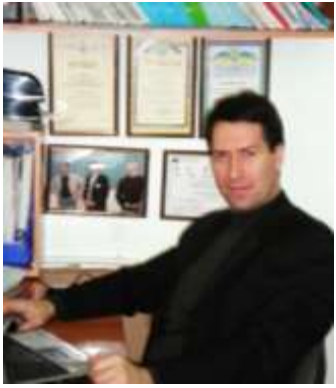

*Majid Mohammad Shafiee*, Ph.D. is an Associate Professor of the Department of Management at University of Isfahan. He graduated from the University of Isfahan in 2013 with PhD in Marketing Management. He is the author of five books and several published papers at national and international levels in refereed journals and conferences. His research interests include competition, financial markets, customer behaviour, and marketing communications.

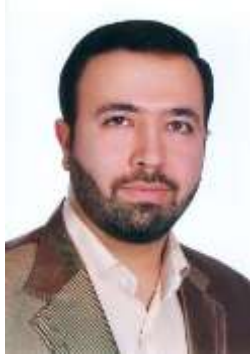

*Malyshenko Vadim Anatolyevich*, Ph.D. is an Associate Professor in Economics of the enterprise and organization of production. His scientific interest is financial standing of enterprise. He is an associate professor of the Department of Economics and Finance of the Institute of Economics and Management of the Humanitarian Pedagogical Academy (branch) of the Federal State Unitary Enterprise «Crimean Federal University named after V.I. Vernadsky». He is the author of 80 publications, including articles indexed in SCOPUS and WoS.

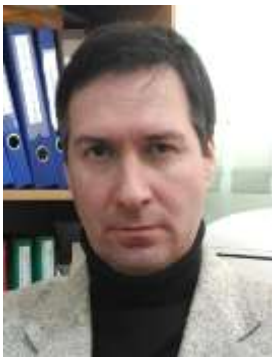

*Anashkina Marina Viktorovna*, M.Sc. Her scientific interest is financial investment, the securities market. In 2016, he graduated from the Institute of Economics and Management of the Humanitarian Pedagogical Academy (branch) of the Federal State Unitary Enterprise "Crimean Federal University named after V. I. Vernadsky" in the specialty "Accounting, analysis and audit". She is the author of 15 publications.

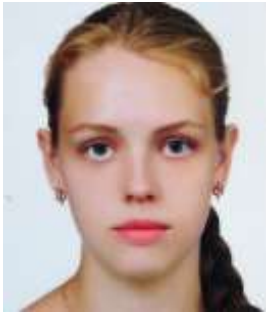

**Abstract**

Developing a system of indicators that reflects the degree to which the securities market fulfils its key functions, is essential to assess the level of its development. In the conditions of asymmetric information it can also provide effective policies for securities market development. This paper is aimed to develop a set of indicators to assess the securities market performance, especially in the asymmetric information context. To this goal, we selected the Russian securities market as a case of asymmetric information context, in comparison with other post-Soviet countries, to investigate its success and failure in fulfilling its key functions. Regarding this, we developed research hypotheses and we conducted a normative research method, based on an ideal model of market functioning that is used as a criterion for testing the hypotheses. The results offer an original scale for assessing the performance of securities market of its functions. The findings also help practitioners with effective policy making in securities market regulation and its development toward its ideal state. The key contribution of this research is in developing a new scale for determining the performance and efficiency of the securities market, based on the conditions of information asymmetry.

**Keywords:** Financial Markets, Information Asymmetry, Market Functions, Securities Market Performance.

**JEL codes:** G10; G18; E44

1. **Introduction**

Assessing the development of securities markets, as well as its efficiency and performance, is of high importance for a countries' economic development. Such an investigation makes it also possible to evaluate the potential growth of an economy, as well as identify the problems of its development. Investigating securities market's indicators can provide mechanisms to assess the dynamics of the market and its nature. These indicators can also be used in economic planning and forecasting by policymakers (Nayak et al., 2019; Zhang and Liu, 2020). However, information asymmetry can also play a crucial role in the securities market and its development. Information asymmetry is an uneven distribution of information between securities market participants. Significant information asymmetry makes the market unattractive for investors. The transparency of such a market is minimal and its participants are not protected from the actions of unscrupulous players, which means they bear increased risks (Ivanitsky and Tatyannikov, 2018; Kuncl, 2019; Li et al., 2019).

Previous research (Acharya and Bengui, 2018; Binner et al., 2018; Celebi and Hönig, 2019; Ellington, 2018; Estrella and Mishkin, 1997; Ghosh and Baduri, 2018, Habimana, 2019; Hashmi and Bhatti, 2019; Rahimi et al. 2020) had tried to assess the state and movement of stock markets, by using a variety of methods and based on both fundamental factors and technical data. However, there are few studies which have focused on securities markets, in investigating its performance by identifying its development measures concurrently in an information asymmetry context. Moreover, there are few studies which focused on the Russian securities market and its development's indices. The emergence of the securities market in Russia and its development entailed the emergence of several problems, therefore further research is necessary. Specifically, the Russian securities market, which is characterized by low liquidity and high potentiality for growth, which make it an emerging market. Some previous researchers who focused on the Russian securities market are Antipina et al. (2018), Bobkov and Shkarina (2019), Chechenova (2017), Fedorova and Gilenko (2008), Khairov and Khasanov (2016), Konovalova and Kuzmina (2019), Lavrenova and Ilyina (2020), Mazaev (2018); Omran (2017), Podgorny (2016, 2017), Rodin and Kadyrov (2018), Solodkaya et al. (2019), Vorobyov and Dikareva (2019), and Yanina et al. (2018). The main methods of analysis in these research are analysing dynamics of average prices, analysing industry stock indices, expert assessments, economic modelling, and analysing changes in trade volume.

Our work is aimed at developing a methodology for evaluating the effectiveness of emerging securities markets, especially in the context of information asymmetry. We introduce a new concept for "functional efficiency of the market". Thus, we want to get an answer to the question of how effectively the securities markets of developing countries work for their economies. The novelty of this approach is that we use monetary aggregates and their ratios, together with other publicly available official indicators, to evaluate securities markets. This makes it possible to develop a universal system of indicators that applies to all countries of the post-Soviet space (since their securities markets are developing). The proposed methodology allows us to determine what place markets occupy in the economy of these countries, and whether there is a potential for their development in order to strengthen the economies.

After reviewing the literature, we will present our materials and methods, which are supplemented with results and discussion.

## 2. Literature review

In this section we review some recent and related research on performance and efficiency of the Russian stock market. Khairov and Khasanov (2016) studied the effectiveness of the Russian stock market. They considered the market from the viewpoint of the performance of its functions. An analysis of key indicators of the securities market indicated an insufficient share of capital in the form of savings of private investors, lack of demand, and lack of interest in the market among the population. The analysis of the share of active clients and individual clients, also reflects the closed nature of the Russian stock market. One way to solve this problem was to popularize the securities market among the population of the Russian Federation, which consequently contributes to its development and reduces dependence on external factors. Antipina et al. (2018) developed indicators to assess the efficiency of the Russian stock market. They also identified the main problems of the Russian securities market as functioning with high volatility and shortage of investment supply in the long period. They further noted that the tools used in the development of regulation measures should be supplemented with mechanisms for assessing stock assets used by institutional investors in the process of analysing the fundamental value of stock assets, which will allow assessing the impact of regulatory measures on market capitalization from the perspective of investors and issuers, as well as making them adjust to ensure consistency with market and fundamental valuation of equity assets. In a similar work, Omran (2017) analysed the effectiveness of the Russian stock market. His paper also examined the information efficiency of the Russian stock market. To this aim, the author used the extended Dickey-Fuller test. It was revealed that a weak form of efficiency is not inherent in the Russian securities market. A similar conclusion was made earlier in the work of Fedorova and Gilenko (2008), where the authors noted that the Russian securities market is ineffective. According to the results obtained, the time series of stock prices are not stationary.

There are a number of features of the domestic market, including a high degree of risks for investors, a general lack of formalization of the market as an institution of collective savings, and a stable source of financing for the national economy. The current state is the result of high inflation rates, an increase in mutual debt of enterprises, and a decline in production and budget deficits. As the main measures of market development, the author notes the need to increase investor confidence, ensure information transparency, and develop a secondary market.

On the other hand, Rodin and Kadyrov (2018) focused on the relationship between the channels of monetary regulation and the efficiency of the stock market. This study raised the question of the importance of the mechanism of money transmission, which ensures the flow of financial capital into various sectors of the economy. It also defines the relationship between the goals and mechanisms of regulation of the national securities market and instruments, which affect the development of stock institutions. In addition, it made an attempt to assess the state of the national financial market on the basis of the M2 aggregate and a number of basic indicators of the securities market, as well as by introducing such an indicator as "the degree of depth of the stock market".

Moreover, Chechenova (2017) studied the functioning of the modern stock market. In his paper, the author notes that the key prospects for the development of the modern stock market, which form the long-term trend of its functioning in the context of globalization, are associated with the movement of global growth to the East, the consolidation of financial institutions, the development of Internet technologies, the segment of derivative securities, the improvement of the market structure, the configuration of the currency basket, and the expansion of the types of financial instruments. At the same time, financial integration, the growing interdependence of stock markets on a global scale, an increase in the risk of rapid spread of crisis phenomena between segments of the global stock market, an increase in the scale and depth of financial crises, instability and market immaturity of countries with economies in transition, give rise to a number of problems in the development of the modern stock market, which implies that a change in regulatory form and approach was proposed.

Yanina et al. (2018) examined the formation and development of the Russian stock market. They analysed how the factors affecting the efficiency of the securities market and the indicators characterizing its profitability was carried out. The estimation of the profitability of the Russian securities market was also given. Konovalova and Kuzmina's (2019) paper is devoted to the study of the emergence, nature, and main characteristics of stock crises. They analysed the financial crises of the Russian stock market. The analysis of the dynamics of prices of financial assets, the values of broad money supply ($M_2X$), the interbank market rate (MIACR), Brent crude oil and the yield of US Treasury bonds were taken as the main indicators. They concluded that a possible prerequisite for unjustified growth of the index levels, is a systematically underestimated asset value, which attracts investors to speculative buybacks, which ultimately leads to a bubble. The model proposed by them makes it possible to identify deviations in any of the areas, and thus, to determine the causes of anomalies arising in the securities market.

In another research, Lavrenova and Ilyina (2020) explored the predictability of the Russian stock market. Their paper examines the main reasons for the low predictability of the Russian stock market. The authors have developed a new approach based on a combination of sectoral and macroeconomic indicators of the functioning of the securities market, the results of which revealed that it has a low sensitivity to changes in such a basic indicator for all markets, such as the price of oil. At the same time, the authors confirmed the preservation of some dependence on changes in the foreign exchange rate. Low market predictability results in high volatility, low investment activity, and a high degree of dependence on the political environment. The influence of these factors does not allow the use of generally accepted models for forecasting. In addition, the authors faced the unavailability of some data required for calculations. Thus, the work focused on the underdevelopment and isolation of the domestic securities market.

Mazaev (2018) also studied the application of in-country macroeconomic methodology for forecasting the dynamics of the Russian stock market. He asserted that the analysis of the macroeconomic level is one of the key stages in the modern mechanism of portfolio management in stock markets with developed economies. In recent years, the popularization of the policy of protectionism, the tightening of currency and commodity wars, have contributed to an increase in the level of independence from external factors and foreign markets in countries with developing economies, which increases the importance of in-country macroeconomic analysis for them. The Russian stock market, thanks to the direction of the fiscal and monetary policies of the regulators, also increases its independence from external factors and, based on the long-term policy of the Government of the Russian Federation and the Bank of Russia, independence will only grow in the future. Due to the long historical dependence of the country's economy on the export of raw material, there is no system of in-country macroeconomic analysis, which, in connection with these structural long-term changes, will improve the quality of portfolio management in the Russian stock market.

Podgorny (2017) investigated the Russian securities market as a social field. This study revealed problems in the perception of the stock market. It noted that the majority of Russians perceive it as "speculative and not performing the function of attracting investment funds". This attitude contradicts the theory of rationality, according to which, as a result of the impact of economic and financial sanctions, the securities market should have become the main platform for attracting and distributing funds for the development of the real economy. A number of cultural and historical factors that prevent the population of the country from understanding the securities market as an affordable tool for earning income from invested funds, are the reasons for this attitude.

Solodkaya et al. (2019) conducted an econometric research by analysing the influence of the structure of the financial market on the economic growth of the Russian Federation. The aim was to study the influence of the ratio of the volume of bank credit and the issue of securities on the rate of economic growth. The indicators characterizing the structure of the financial market were the volumes of bank lending (with the allocation of loans to individuals and organizations) and the total market capitalization of shares traded on the Moscow Exchange. It is shown that economic growth largely depends on the development of bank credit and, to a lesser extent, on the growth of the market capitalization of shares.

Although studying the securities market and the economy, based on the comparison of monetary aggregates as well as their comparison with various fundamental indicators (such as GDP, industry indices, etc.) is widely used by previous research, they are not used directly to analyse the performance and dynamics of securities markets. Table 1 summarizes the most important indicators we used after reviewing the literature.

**Table 1. Characteristics of the most significant indicators in the literature\***

| Authors | Indicator/ Methodology/ Methods | Goal/ Advantage | Limitation/ Disadvantage |
|---|---|---|---|
| Antipina et al. (2018) | Application of the APV asset valuation model. | Assessing the impact of government regulatory measures on the capitalization of the securities market from the perspective of investors and issuers. Evaluating the effectiveness of the securities market. | The complexity of getting some data and making calculations. The indicator is mainly needed to improve government regulation measures. |
| Estrella and Mishkin (1997) | Multivariate analysis. Optimal policy and the McCallum rule. | They are used for the purpose of assessing monetary policy. | There are certain limitations in the use of monetary aggregates for providing information about the effectiveness of monetary policy. These indicators also do not allow us to answer the question about the effectiveness of the functioning of the securities market |
| Hashmi and Bhatti (2019) | GDP-weighted growth rates method. Currency equivalent (CE) method. | They are based on the analysis of GDP, monetary aggregates, and the division index. Assessing the global liquidity and developing certain measures to improve the monetary position. | Although they are not aimed at the securities market, these indicators are interesting for our research, as they allow us to confirm the relationship between the indicators we have selected. |
| Mazaev (2018) | Leading and confirming indicators. | Leading macroeconomic indicators allow us to make a qualitative forecast of the dynamics of the country's economy and the stock market for the next 3-6 months. The confirmatory ones allow us to check the correctness of the results obtained. | They do not allow us to assess the role of the securities market in the development of the country's economy. It is aimed at studying and forecasting the Russian economy and, as a result, the stock market. However, the relationship between the degree of development of the economy and the securities market for developing countries is not unambiguous. |

*\* Compiled by the authors*

## 3. Materials and methods

### 3.1. Case study: the Russian stock market as an emerging securities market

We considered the Russian securities market for our case study, since it is at the stage of formation and can be inspected as an accurate sample of asymmetric information context. Due to the weak development of market relations, high inflation, and frequent socio-political instability in emerging economies, the securities market is not able to fully perform its main functions. An increase in the degree of reliability and confidence in the securities market depends mainly on an increase in the level of its organization and strengthening of state control over it. It is also considered as an emerging financial market, which is part of a developing market. Developing markets are usually called markets, whose capitalization is less than 1/10 of the world level[1]. Due to the rapid growth of emerging markets, a monthly trading volume of at least $ 2 billion and the number of registered companies of at least 100, have recently become a criterion for classifying them as developing markets (Mityai, 2019). An emerging market can also be measured by some liquidity in local debt markets, as well as by the existence of some forms of market relations and regulators. The main characteristics of the securities market in Russia in comparison with other post-Soviet countries are shown in Figure 1.

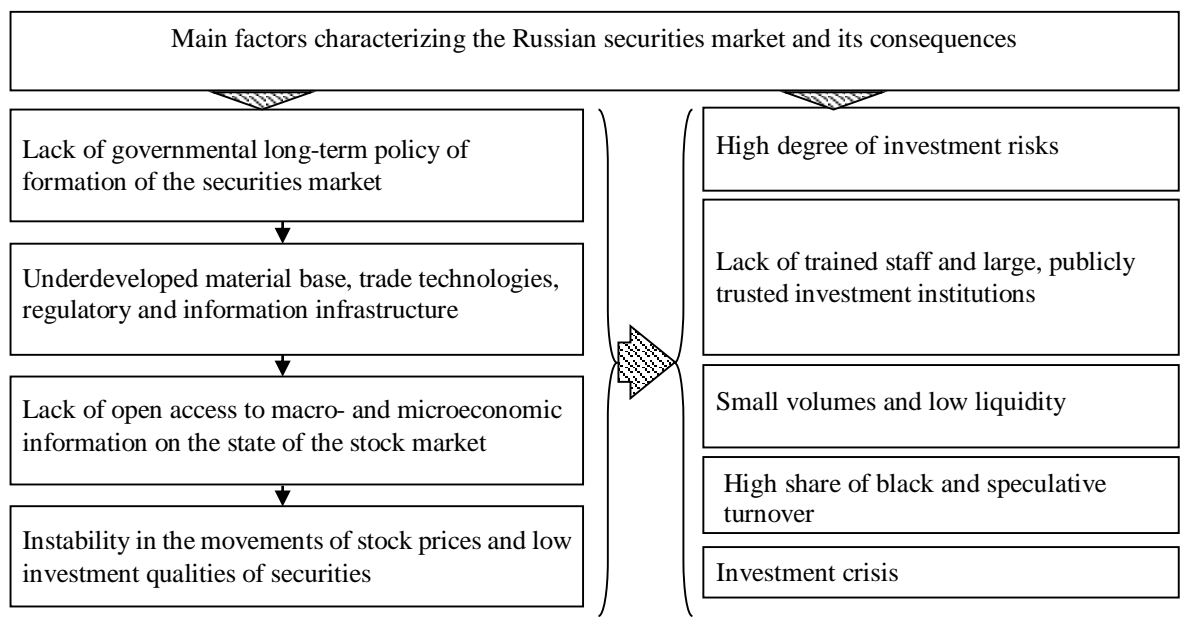

**Figure 1. Key characteristics of emerging securities markets\***

*\* Compiled by the authors*

As an economic mechanism, the securities market provides the distribution of capital, across sectors of the economy. Previous works (Antipina et al., 2018; Chechenova, 2017; Khairov and Khasanov, 2016) suggest several functions for financial markets, and specifically for stock markets. After reviewing the most important functions of the stock market, we developed a framework for conceptualizing our intended variables, so that in the next stage, we can test the main hypotheses regarding the main research question. Figure 2 illustrates our research framework, which conveys key functions in securities markets.

---

[1] https://en.wikipedia.org/wiki/Emerging_market

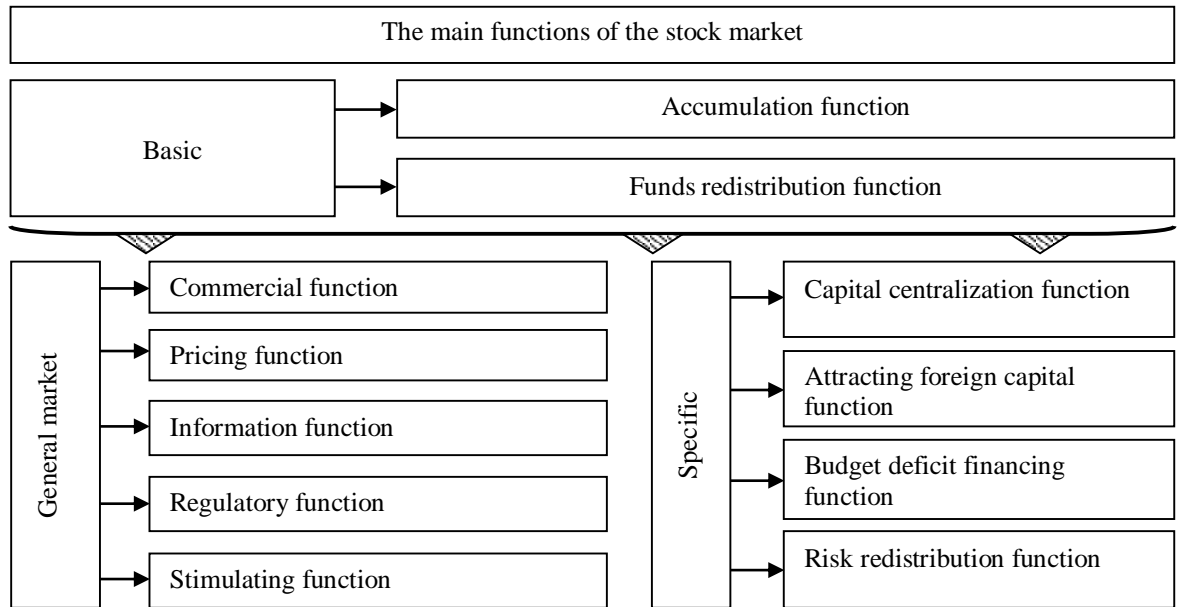

**Figure 2. The main functions of the stock market\***

*\* Compiled by the authors*

As illustrated in Figure 2, *Basic functions* reflect the essence of the securities market. These functions include accumulation and accumulation, as well as distribution of funds. They are also measures of the market efficiency and the degree of its significance for the development of a country's economy. The performance of these functions by the securities market is more important at the macroeconomic level.

*General market functions* are commercial, pricing, information, regulatory, stimulating, which are important mostly at the meso- and micro-level, since their implementation ensures the redistribution of funds and information between its participants, setting prices for financial market instruments, making a profit, and ensuring self-regulation of the securities market.

*Specific market functions* are those functions that are specific exclusively to the securities market. These include centralizing of capital, financing the budget deficit, redistributing of risks, and attracting foreign capital.

The effective activity of a securities market contributes to the development of production, an increase in turnover, the transformation of cash receipts from individuals and legal entities into investment. It is also a source of recovery of fixed capital. Thus, the purpose of the function of the securities market is to form a mechanism for attracting investment to the economy, and to its real sector, for the renewal of industry.

Regarding the key functions of securities markets, our main research question is how well is the Russian securities market functioning? To answer this question, it is necessary to discover and analyse key indicators that reflect the movement of money supply, so that we can estimate the performance of the securities market. In this regard, this study proposes a methodology based on the proportions of monetary aggregates and also on specific indicators of the securities market. Since the purpose of this study is to assess the efficiency of the stock market, the authors introduce the concept of functional efficiency, which has rarely been considered in past works, and thus shapes one of our key contributions. Based on our definition, *the functional efficiency of the securities market* is a quantitative assessment of the quality of securities market performance with its main functions, which are illustrated in Figure 2. The concept focuses on assessing the quality and completeness of the market performance with its main functions. The concept can also be used as an indicator to estimate the development of an economy, from the securities market perspective.

Regarding the above-mentioned framework and our underlying theory, we formulate our research hypotheses as below:

*$H_1$: The Russian securities market performs the accumulation function.*

*$H_2$: The Russian securities market performs the redistribution function.*

The selection of these hypotheses, out of other possible functions, is justified by the fact that the other functions are performed by the market spontaneously, when these two basic functions are not violated. Since our results show that these two hypotheses are not supported, we then remove other hypothesized functions of our research and suggest them to future studies. In our work, we thus concentrate on the performance of two main functions: accumulation and distribution. The supremacy of these functions stems from the main goal of the securities market, which is to mediate in the movement of temporarily free funds to issuers to ensure the development of the economy.

### *3.2. The analysis of indicators and methodology*

The proposed methodology for studying the performance of the securities market with its functions is mainly based on the method for assessing the movement of money supply at the macroeconomic level, which is possible when analysing monetary aggregates. To test our hypotheses, it is thus necessary to analyse the general indicators of the securities market and changes in the indicators of the money supply. The need for developing such a methodology is due to the fact that, the securities markets of developing countries cannot be analysed using merely the methods established for developed markets. There are significant differences. Moreover, the existing methodologies for emerging markets have some limitations and disadvantages. In addition, there is a problem with the unavailability of data that is necessary for evaluating the market regarding previous methods. That is why we based our method on the money supply indicators, which can reflect the movement of funds and are available through official data.

The money supply indicators used by most economists are variants of what we call them $M_1$, $M_2$, and $M_3$ aggregates. The structure and formulas for calculating monetary aggregates of the post-Soviet countries are presented in Table 2.

**Table 2. Ratio of monetary aggregates***

| Monetary aggregate / Country | $M_0$ | $M_1$ | $M_2$ | $M_{2X}$ | $M_3$ |
|---|---|---|---|---|---|
| Russia, Ukraine, Moldova, Uzbekistan, Tajikistan, Turkmenistan, Georgia, Azerbaijan, Armenia, Lithuania, Estonia, Latvian | Banknotes and coins in circulation (the most liquid monetary aggregate, cash outside of deposit banks) | $M_0$+ transferable deposits, representing the balances funds of non-bank credit and commercial and non-commercial organizations, and household deposits in banks on demand | $M_1$+ term deposits in banks, savings deposits, as well as compensation | | $M_2$+ government loan certificates and bonds |
| Kazakhstan | | | | | $M_2$ + term and savings deposits in foreign currency |
| Belarus | | | | $M_2$ + funds in securities (other than shares) of non-bank credit and financial organizations, commercial and non-commercial organizations, individual entrepreneurs and residents of the Republic of Belarus | $M_{2x}$ + transferable and term deposits in foreign currency, as well as securities (other than shares) in foreign currency, deposits in precious metals of non-bank credit and financial, commercial and non-commercial organizations, and individual entrepreneurs |
| Kyrgyzstan | | | $M_1$ + term deposits of the population in commercial banks, plus short-term government securities | $M_2$ + all types of deposits in foreign currency, converted at the official exchange rate of the national currency | $M_2$+ government loan certificates and bonds |

*\* Compiled by the authors*

The money supply of the post-Soviet countries has approximately the same structure. As can be seen from table 2, the set of aggregates that characterize it, as well as the formulas for calculating them, are the same in most countries. The exceptions are Kazakhstan, the Republic of Belarus, and Kyrgyzstan. These countries differ in the calculation of the indicators $M_2$ (Kyrgyzstan), $M_3$ (Kazakhstan, the Republic of Belarus).

In our case, Russia, the $M_2$ aggregate is used as the most universal indicator of the money supply. It is also noticeable that the stock and money markets are different sets of financial instruments and institutions, but they have a close relationship. The change in the proportions of the aggregates that characterize the money market acts as an indicator of the efficiency of the functioning of the securities market.

The circulation of the money supply of the country is based on maintaining the equilibrium of all monetary aggregates. It occurs, when the unit $M_1$ is larger than the most illiquid ($M_0$), or in the case when $M_2 + M_3 > M_1$. Money capital then goes from cash to non-cash circulation. This indicates that there is no significant increase in prices in the country. The existing methods do not allow assessing the performance of the main functions of the market, i.e., accumulation and redistribution of capital, but assessment is possible through a system of macroeconomic indicators.

The official statistical websites of the countries under review publish data on the $M_0$, $M_1$ and $M_2$ aggregates. However, not all countries officially calculate the $M_3$ aggregate, which usually includes stock market data. In this regard, a securities market assessment of its main functions cannot be made on the basis of a simple ratio of the available data on aggregates. Thus, the following statistical data can be added: monetary aggregates and monetary base, GDP (Gross Domestic Product) values, data on money income and savings of the population, open data on the securities market, and inflation rates. All of these indicators are published on the official websites of the state statistics of the countries.

Moving on to assessing the performance of the securities market by its functions, it is necessary to determine the conditions under which the functions will be considered performed. The decrease in the cash supply in the country indicates that citizens are actively using the services of the banking sector - the movement of funds is turning toward a non-cash form. The money supply, thus, circulates in various sectors of the economy, including through the securities market. In the money supply, the savings of the population are also taken into account. The aggregates do not reflect the movement of savings across certain sectors, however, comparing them with different indicators, it becomes possible. Thus, we can declare the performance of the accumulation function by the securities market if:

*- With an increase in the share of non-cash funds, an increase in the volume of the securities market is observed;*

*- With a decrease in the volume of household savings, an increase in the volume of the securities market is observed;*

*- With a decrease in the volume of savings of the population, there is an increase in the $M_3$ aggregate, as well as:*

*- If the volume of savings of the population decreases or does not change with an increase in market volumes, the value of $M_0$ decreases or is constant;*

*- If, with an increase in market volumes, $M_0$ is constant, and $M_1$ and $M_2$ increase.*

These conditions can be represented by Formula 1:

$$\text{Accum.} = \begin{cases} \mu_0 => \uparrow \\ Q_{sm} => \uparrow \end{cases} \wedge \begin{cases} S_{pop.} => \downarrow \\ Q_{sm} => \uparrow \end{cases} \wedge \begin{cases} S_{pop.} => \downarrow \\ M_3 => \uparrow \end{cases}, \qquad (1)$$

Where:

*Accum. – accumulation function;*

*$\mu_0$ – aggregate money multiplier $M_2$ and $M_0$;*

*$M_3$ – aggregate $M_3$;*

*$Q_{sm}$ – volume of securities and other assets in the accounts of participants in the market;*

*$S_{pop.}$ – Population savings.*

In this case, the calculation of indicators is carried out by Formula 2:

$$\mu_0 = \frac{M_2}{M_0}, \qquad (2)$$

Where:

*$M_2$ – aggregate value $M_2$;*

*$M_0$ – aggregate value $M_0$.*

The calculation of the indicated money multiplier includes the values of the non-monetary base, as is customary, but of $M_0$ aggregate. This choice is due to the presence in the value of the monetary base of such components as cash in mandatory reserves and cash of commercial banks in correspondent accounts of the Banks. Since the indicated values, due to the specifics of their own functions, cannot be placed on the stock markets, they are not included in the calculation. At the same time, to assess the functions of the securities market, it is important to take into account the dynamics and structure of the money supply, in particular - the change in the share of non-cash funds in circulation, since this value reflects the qualitative changes in the economy. The predominance of cash in circulation is a negative factor, while the transition to a non-cash form (accordingly, an increase in $M_2$) indicates a high degree of economic development and its qualitative transformation.

Accordingly, by comparing the presented multiplier and the dynamics of the securities market volumes, it becomes possible to assess its performance of the accumulation function. So, if there is an increase in the non-cash supply, but at the same time there is no growth in the stock market, or it is insignificant, we can say that the population's funds remain in the accounts of the banking system (either remain unclaimed, or are placed in deposit accounts). Accordingly, the money supply does not move to the securities market accounts, which indicates a failure to perform the function. Thus, we can talk about the execution of a function, if the condition presented in the first part of the formula is met.

In this case, the following condition should be highlighted: the market dynamics can be considered positive if $\Delta Q_{sm} > \pi$ (where $\pi$ is the inflation rate), and insignificant if $\Delta Q_{sm} \leq \pi + 10\%$ (we have assumed a significant level of market growth of 10 percent above the inflation rate). Thus, the calculation of the volume of savings of the population was made according to the following formula (formula 3):

$$S_{pop.} = M_0 + (M_2 - M_1) \qquad (3)$$

Comparison of the dynamics of the market volume and the dynamics of the population's savings makes it possible to track the movement of the money supply (the second part of the formula). Decrease of the *Spop* indicator, with an increase of the *Qsm* indicator, indicates that the population is investing money in securities, that is, the money supply is moving to the accounts of securities market participants. If the volume of savings increases, and the volume of the securities market decreases, the opposite situation is observed. An increase in the volume of savings with a constant or insignificant change in the volume of the securities market indicates that the securities market does not perform the function of accumulation. In this case, the condition presented above remains valid.

Next, we present the calculation of the $M_3$ unit, which is not published by the statistical services of some countries (for example, Russia). The indicator usually includes certificates and domestic bonds, so the indicator is broad and reflects the total money supply of the state. Thus, if there is no data, it can be calculated based on the available data on the volume of trading in the bond market. Then you need to use the value of $M_3$ using Formula 4:

$$M_3 = M_2 + Q_{sm.} \qquad (4)$$

Thus, when comparing changes between $M_3$ and $S_{pop.}$ it is possible to determine whether the money supply has been transferred to the bond market, which is part of the securities market. As for the previous components, the dynamics of $M_3$ can be considered positive if $\Delta M_3 > \pi$, and insignificant if $\Delta M_3 \leq \pi + 10\%$. Then the condition under which the accumulation function is fulfilled can be represented as follows (Formula 5):

$$\text{Accum.} = \begin{cases} \mu_0 => \uparrow \\ Q_{sm} => \uparrow \\ S_{pop.} => \downarrow \\ M_3 => \uparrow \end{cases} \text{By} \begin{cases} \Delta Q_{sm} > \pi \\ \Delta S_{pop.} > \pi \\ \Delta M_3 > \pi \end{cases} \qquad (5)$$

We then assessed the performance of the funds reallocation function. The redistribution function provides funding for various areas of the economy, which results in money being transformed into real investment flows that can reach maximum values in the most promising sectors. As a result of the distribution of investment flows, the uniform development of various industries is carried out, which ultimately is reflected in the GDP indicator. The effective implementation of the redistribution function by the securities market allows industries to increase the volume and quality of their products. The growth of production, in turn, has an impact on the volume of monetary aggregates - there can be a movement of the money supply from one aggregate to another, due to the growth or decrease of one or another of its components, as well as an increase in the money supply as a whole.

Based on the definition of the function of redistribution of funds, we can consider the relationship of three indicators: monetary aggregates, production volumes and the volume of GDP. The main task of the creation and existence of the securities market is the redistribution of funds in such a way that all sectors of the economy have the opportunity for development through high-quality transformation of funds, and the introduction of the latest technologies. We have assumed the following condition (Formula 6):

$$\text{Redistr} = \begin{cases} \Delta \frac{Q_{tr.}}{M_3} => \uparrow \\ \Delta \frac{c}{Q_{sm}} => \uparrow \end{cases} \quad \text{By} \begin{cases} \Delta Q_{sm} > \pi \\ \Delta GDP > \pi \end{cases} \qquad (6)$$

Where:

*Redistr – redistribution function;*

*$Q_{tr.}$ - trading volume;*

*GDP - Gross domestic product.*

In the final step, we checked the reliability of the results obtained by the indicators that were developed by the authors. To assess the degree of performance of the function and the level of development of the market as a whole, it is necessary to analyse such values as the turnover of funds in the stock market and the maximum limiting volume of the securities market under given economic conditions. A special condition in solving this problem is the use of publicly available official information in the calculations. Under the given conditions, we can use the indicators developed to assess the performance of the previous function, derived from monetary aggregates. The following indicators are presented for calculation (Formula 7 to 10).

1. The turnover ratio of the securities market (Formula 7):

$$k_{tur.} = \frac{Q_{tr.}}{Q_{sm}} / 100 \qquad (7)$$

Where:

$k_{tur.}$ – *Securities market turnover ratio.*

The indicator represents the quotient of the trading volume in the securities market and the volume of the securities market directly. Thus, it is possible to estimate how many turnovers make 1 Ruble of market participants' funds per year. The indicator is considered both individually (in terms of dynamics) and in conjunction with other indicators. When studying dynamics, its positive value will reflect an increase in the turnover of market assets, i.e. an increase in the activity of its participants. Therefore, this indicator can also be called the activity coefficient. The higher its value, the more efficient the market will work in the given time frame under study.

2. Maximum market limit for given economic conditions (Formula 8):

$$\lim\nolimits_{max} = S_{pop.} + Q_{sm} + (M_1 - M_0) \qquad (8)$$

Where:

$lim_{max}$ – *Maximum market limit under given economic conditions.*

Since the main objective of this study was to assess the degree to which the market performs its functions, it is necessary to understand what can be considered as its limit in the current economic situation. In other words, the function of this indicator is to reflect the maximum possible market volumes, when its efficiency will be 100% with the money supply available in the country.

The indicator is based on the values of monetary aggregates, the volume of savings of the population, previously calculated on their basis, and the volume of the securities market. The market limit is the potential sum of all money resources of the population, which hypothetically can take part in the circulation on the securities market. These are funds, not only in cash, and in the form of term deposits in bank accounts, or directly in market accounts, but also funds in bank accounts on demand. The value of the indicator is not constant, and changes depend on the economic situation and changes in the proportions of monetary aggregates, which allows using the following indicator to objectively assess the level of development and the degree of market functioning efficiency. The disadvantage of this indicator is that it does not take into account possible receipts from foreign participants, and thus only measures the domestic potential. But, the purpose of this study is to precisely determine the availability of the market for the population of the country, and accordingly, to study its functional efficiency, specifically for the economies of the studied countries.

3. Indicator of the efficiency of the securities market (Formula 9):

$$FE_{sm} = \frac{Q_{tr.}}{lim_{max} * k_{tur.}} \qquad (9)$$

Where:

*$FE_{sm}$ – Functional efficiency of the securities market.*

The values of the indicator as a percentage allow us to assess the efficiency (power) of the market with the available opportunities. The higher the value of the indicator, the more efficiently the market functions.

4. The calculation of efficiency allows us to determine the market potentiality (Formula 10):

$$SMP = 100 - FE_{sm} \qquad (10)$$

Where:

*SMP – Securities market potential.*

This indicator is the inverse of the efficiency indicator, and reflects the share of funds that are potentially involved in the turnover of the securities market. The higher the value, the lower the market efficiency.

In order to make sure that the proposed methodology can be applied to all countries of the post-Soviet space, we conducted an analysis of the compliance of the necessary indicators. As mentioned earlier, Kazakhstan and Kyrgyzstan have some differences in the calculation of monetary aggregates. Therefore, we propose the following correspondence matrix (Table 3). Using the formulas presented in the table, the required indicator can be found.

**Table 3. Correspondence matrix for the selection of indicators of the post-Soviet countries**

| Country<br>Indicator | *Russia* | *Ukraine* | *Belarus* | *Moldova* | *Uzbekistan* | *Kazakhstan* | *Tajikistan* | *Turkmenistan* | *Kyrgyzstan* | *Georgia* | *Azerbaijan* | *Armenia* | *Lithuania* | *Latvian* | *Estonia* |
|---|---|---|---|---|---|---|---|---|---|---|---|---|---|---|---|
| $M_0$ | + | + | + | + | + | + | + | + | + | + | + | + | + | + | + |
| $M_1$ | + | + | + | + | + | + | + | + | + | + | + | + | + | + | + |
| $M_2$ | + | + | + | + | + | + | + | + | $M_2$-short-term government securities | + | + | + | + | + | + |
| $M_3$ | + | + | + | + | + | $M_3$+ government securities | + | + | $M_2$ + government securities | + | + | + | + | + | + |
| $Q_{sm}$ | + | + | + | + | + | + | + | + | + | + | + | + | + | + | + |
| $\pi$ | + | + | + | + | + | + | + | + | + | + | + | + | + | + | + |
| $Q_{tr.}$ | + | + | + | + | + | + | + | + | + | + | + | + | + | + | + |
| GDP | + | + | + | + | + | + | + | + | + | + | + | + | + | + | + |

*\* Compiled by the authors*

## 4. Results and discussion

In order to test our proposed methodology, we chose data from Russia, since it is the country with the largest GDP and securities market among the countries of the post-Soviet space. We chose a standard set of indicators that is usually used to analyse the performance of the securities market. The main indicators and their values are presented in Table 4.

**Table 4. Main indicators of the securities market and their values**[*, **, ***]

| *Indices* | *2012* | *2013* | *2014* | *2015* | *2016* | *2017* | *2018* |
|---|---|---|---|---|---|---|---|
| No. of professional participants in the market | 1235 | 1149 | 1079 | 875 | 659 | 599 | 520 |
| Trading volume on stock exchanges, including: | 24 132 | 24 025 | 20 887 | 20 556 | 23 893 | 35 414 | 40 671 |
| Stock market | 11 647 | 8 707 | 10 283 | 9 398 | 9 277 | 9 185 | 10 830 |
| Bond market | 12 485 | 15 319 | 10 605 | 11 159 | 14 616 | 26 228 | 29 841 |
| Money market | 178 674 | 220 708 | 204 375 | 213 786 | 333 883 | 377 141 | 364 216 |
| Currency market | 116 980 | 156 016 | 228 546 | 310 837 | 329 954 | 347 671 | 348 368 |
| Derivatives market, including: | 49 969 | 48 605 | 61 316 | 93 713 | 115 271 | 84 497 | 89 263 |
| Futures | 46 760 | 44 588 | 55 566 | 90 231 | 109 489 | 77 624 | 82 397 |
| Options | 3 209 | 4 017 | 5 749 | 3 482 | 5 782 | 6 873 | 6 866 |

*\* The Central Bank of the Russian Federation (URL: http://www.cbr.ru), \*\* at the end of the year, \*\*\* billion Rubles*

The increase of indicators is characterized by the influence of various factors, including inflation. A decrease in the number of professional participants (a decrease by 715 units in 2018 compared to 2012), with a simultaneous increase in market volumes, is indirect evidence of its consolidation. This is also confirmed by a decrease in the volume of trading in shares (-7%). A significant increase applies to the foreign exchange and money markets (by 2018, the indicators increased by 197% and 103%, respectively). By 2018, the volume of issue of debt securities is significantly increased (by 149%) (See Table 5). In the structure of securities, the volume of short-term issues is increasing, however, long-term issues still prevail (93% of the total volume in 2018). The issue of debt securities in foreign currency is growing 13 fold compared to 2012.

**Table 5. Number and volume of registered securities issues in the Russian Federation**[*,\*\*,\*\*\*]

| Registered issues of securities | 2012 | 2013 | 2014 | 2015 | 2016 | 2017 | 2018 |
|---|---|---|---|---|---|---|---|
| Total debt securities issued | 8 516 503 | 9 779 945 | 12 237 465 | 13 600 298 | 15 616 388 | 19 144 403 | 21 220 826 |
| Short term | 13 000 | 2 000 | - | - | 4 086 | 399 889 | 1 396 218 |
| Long-term | 8 503 503 | 9 777 945 | 12 237 465 | 13 600 298 | 15 612 302 | 18 744 514 | 19 824 608 |

| Debt securities issued in Rubles | 8 472 169 | 9 719 080 | 12 082 211 | 13 293 486 | 15 122 429 | 18 666 788 | 20 630 377 |
|---|---|---|---|---|---|---|---|
| Short term | 13 000 | 2 000 | - | - | 4 086 | 399 889 | 1 396 218 |
| Long-term | 8 459 169 | 9 717 080 | 12 082 211 | 13 293 486 | 15 118 343 | 18 266 899 | 19 234 159 |
| Volume of debt securities issued in foreign currency | 44 334 | 60 865 | 155 254 | 306 812 | 493 959 | 477 615 | 590 449 |
| Short term | - | - | - | - | - | - | - |
| Long-term | 44 334 | 60 865 | 155 254 | 306 812 | 493 959 | 477 615 | 590 449 |

*\* The Central Bank of the Russian Federation (URL: http://www.cbr.ru), \*\* at the end of the year, \*\*\* million Rubles*

Regarding the function of redistribution of funds, one should pay attention to the dynamics of the market for sectorial stock indices (See Figure 3). The study of market dynamics and its structure makes it possible to characterize the efficiency of the redistribution of funds through the system of the securities market in certain sectors of the economy. Sectorial indices during 2015-2018 changed in different directions (Figure 3).

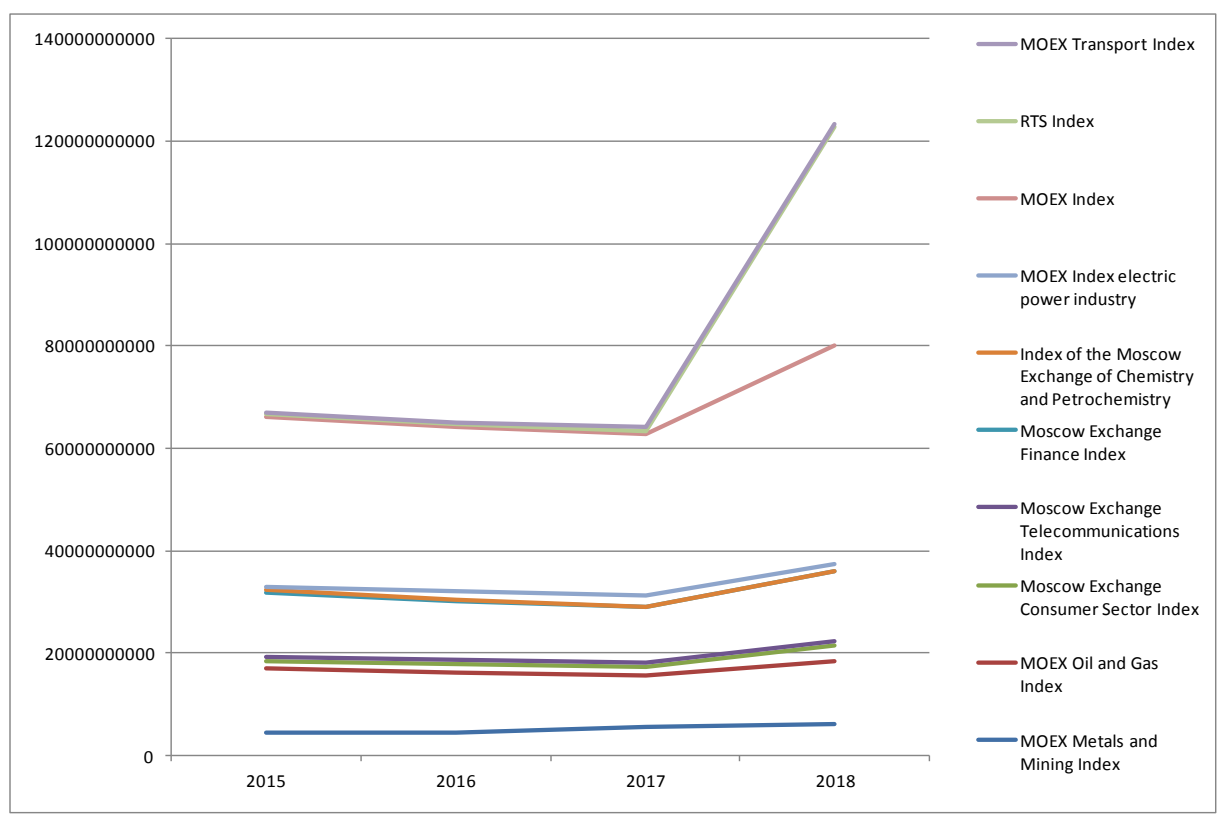

**Figure 3. Average annual values of the volumes of indices**[*,**,***]

*\* The Central Bank of the Russian Federation (URL: http://www.cbr.ru), \*\* at the end of the year, \*\*\* Rubles*

The indices of "Metals and Mining" and "Consumer Sector" demonstrated stable growth. The volumes of the Oil and Gas and Finance indices changed insignificantly. As can be seen in the graph, these indices have the highest values. The distribution presented in the graph is an insignificant characteristic of the efficient operation in the Russian securities market. As noted earlier, to assess the effectiveness of securities market functions performance, a standard analysis of indicators is not enough.

In order to analyse the securities market based on the proposed methodology, data on the indicators of money circulation were gathered. As shown in Figure 4, the volume of money supply increases during the time, while the share of cash tends to decrease. The circulation rate of the money supply remains stable.

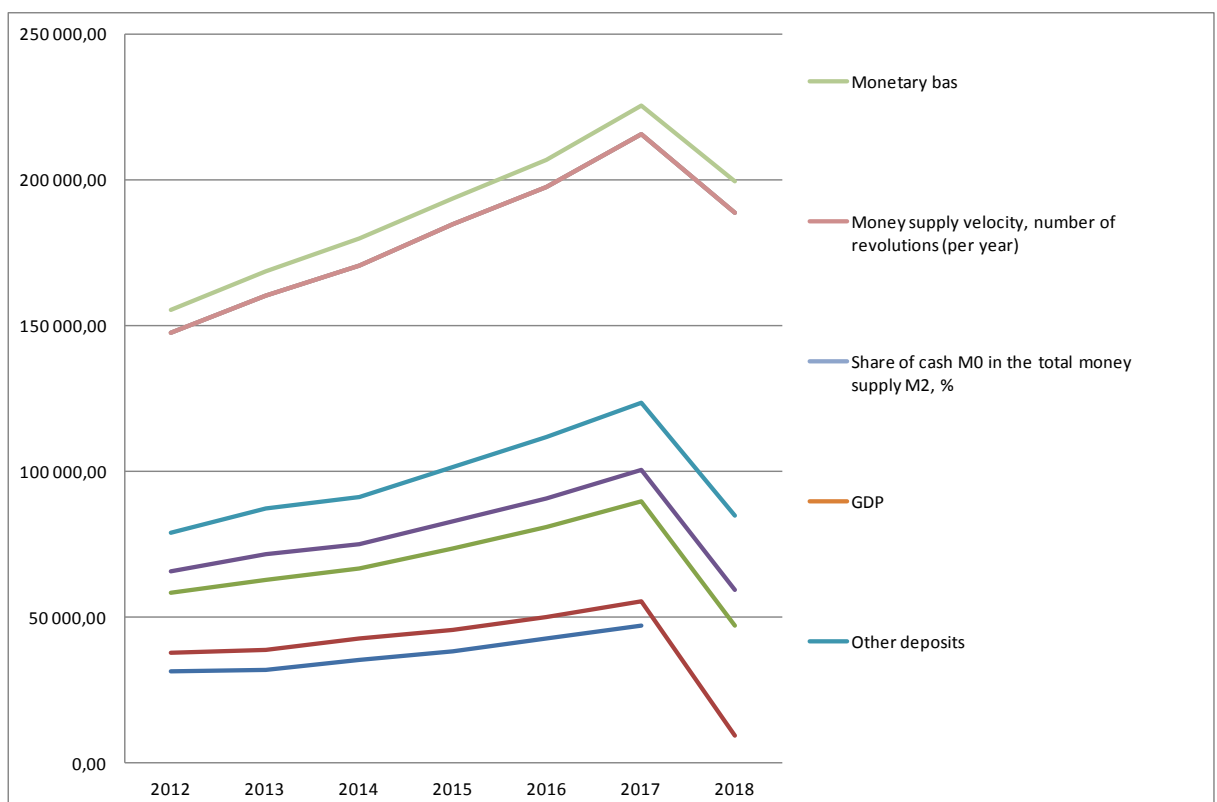

**Figure 4. Main indicators of money circulation**[*,**,***]

*The Central Bank of the Russian Federation (URL: http://www.cbr.ru), ** at the end of the year, *** billion Rubles*

Table 6 summarizes our results, which are the output of carrying out the calculation of the necessary indicators. The calculation was carried out by using data from official sources of the last two years (2017 and 2018) (regular reports of the Central Bank of the Russian Federation).

**Table 6. Indicators for assessing the performance of the securities market accumulation function**[*,**,***]

| Index \ Year | 2017 | 2018 | Change in indicator to the previous year | | Inflation rate (π), % |
|---|---|---|---|---|---|
| $M_0$ | 7 860,6 | 8 762,8 | 11,4 | ↑ | |
| $M_1$ | 17 787,2 | 20 025,9 | 12,5 | ↑ | |
| $M_2$ | 39 085,3 | 43 384,3 | 10,9 | ↑ | |
| $Q_{sm}$ | 912,5 | 1 070,9 | 17,3 | ↑ | 4,27 |
| $M_3$ | *39 997,8* | *44 455,2* | 11,1 | ↑ | |
| $S_{pop.}$ | *29 158,7* | *32 121,2* | 10,1 | ↑ | |
| $\mu_0$ (coefficient) | *4,97* | *4,95* | -0,4 | ↓ | - |

*\* The Central Bank of the Russian Federation (URL: http://www.cbr.ru), ** at the end of the year, *** billion Rubles*

$M_3$ is required to calculate the presented indicators. By substituting the obtained results into the previously derived condition, we have Formula 11 and 12:

$$\text{Accum.} = \begin{cases} \mu_0 => \downarrow \\ Q_{sm} => \uparrow \end{cases} \wedge \begin{cases} S_{pop.} => \uparrow \\ Q_{sm} => \uparrow \end{cases} \wedge \begin{cases} S_{pop.} => \uparrow \\ M_3 => \uparrow \end{cases} \quad (11)$$

Or:

$$\text{Accum.} = \begin{cases} \mu_0 => \downarrow \\ Q_{sm} => \uparrow \\ S_{pop.} => \uparrow \\ M_3 => \uparrow \end{cases} \text{By} \begin{cases} \Delta Q_{sm} > \pi \\ \Delta S_{pop.} > \pi \\ \Delta M_3 > \pi \end{cases} \quad (12)$$

Thus, none of the conditions have been met, therefore it can be concluded that the accumulation function is not performed by the Russian securities market. Thus, Hypothesis $H_1$ is not supported.

We then summarized the necessary data after calculating the indicators (Table 7). As shown in Table 7, the selected indicators have a tendency, which negatively characterizes the fulfilment by the securities market on the function of redistributing funds. With the growth of GDP and $M_3$ indicators, there are negative dynamics on the share of investments in fixed assets and the share of products of high-tech and science-intensive industries. The growth of the inventive activity ratio is insignificant.

**Table 7. Indicators for assessing the performance of the securities market redistribution function**[*,**,***]

| Year / Index | 2017 | 2018 | Change in indicator to the previous year | | Inflation rate ($\pi$), % |
|---|---|---|---|---|---|
| GDP | 92 089,3 | 103 626,6 | 12,5 | ↑ | |
| $M_3$ | 39 997,8 | 44 455,2 | 11,1 | ↑ | 4,27 |
| $Q_{tr.}$ | 887 569 | 861 119 | -3,1 | ↓ | |
| $\omega_{s.i.}$ (% to GDP) | 21,6 | 21,1 | -2,3 | ↓ | - |
| $\omega_{I.}$ (% to GDP) | 21,5 | 20,7 | -3,7 | ↓ | - |
| $K_{inv.}$ | 1,55 | 1,70 | 9,6 | ↑ | - |
| $Q_{tr.}/M_3$ (%) | 22,1 | 19,3 | -12,6 | ↓ | - |

*\* The Central Bank of the Russian Federation (URL: http://www.cbr.ru), \*\* at the end of the year, \*\*\* billion Rubles*

The trading volume decreased by 3.1%, by 2018 with an inflation rate of 4.27%, based on which it can be concluded that the market as a whole had a negative trend. In other words, during 2017-2018, the Russian securities market did not develop. The negative values of the GDP components, which are important in assessing the securities market, indicate that during the study period, the function of redistributing funds was not performed by it. We substitute the obtained values into the derived condition (Formula 13).

$$\text{Redistr} = \begin{cases} \Delta \frac{Q_{tr.}}{M_3} => \downarrow \\ \Delta \omega_{s.i.} => \downarrow \\ \Delta \omega_I => \downarrow \\ \Delta k_{inv..} => \uparrow \end{cases} \text{By} \begin{cases} \Delta Q_{tr.} < \pi \\ \Delta DGP > \pi \end{cases} \quad (13)$$

As can be seen, the condition is not met, which allows us to assume that the Russian securities market does not perform the redistribution function. Thus, Hypothesis $H_2$ is not supported.

According to the above-mentioned formulas, we calculated the indicators, which are presented in Table 8. According to the results, it is clear that the efficiency of the Russian securities market in 2018 was about 2.4%, that is, 97.6% of potential resources have not been attracted. This allows us to conclude that the market is ineffective as a financial policy instrument. The market acts as a platform for speculative transactions, as evidenced by the growth in trading volumes in the foreign exchange and money markets, as well as a decrease in the number of professional market participants, and the volume of trading in shares.

**Table 8. Calculation of performance indicators of the Russian securities market**

| Year / Index | 2017 | 2018 | Change in indicator to the previous year | |
|---|---|---|---|---|
| $M_0$ | 7 860,6 | 8 762,8 | 11,4 | ↑ |
| $M_1$ | 17 787,2 | 20 025,9 | 12,5 | ↑ |
| $M_2$ | 39 085,3 | 43 384,3 | 10,9 | ↑ |
| $Q_{sm}$ | 912,5 | 1 070,9 | 17,3 | ↑ |
| $S_{pop.}$ | 29 158,7 | 32 121,2 | 10,1 | ↑ |
| $Q_{tr.}$ | 887 569 | 861 119 | -3,1 | ↓ |
| $K_{tur.}$ | 9,7 (972,6) | 8,04 (804,1) | -11,7 | ↓ |
| $lim_{max}$ | 39 997,8 | 44 455,2 | 11,1 | ↑ |
| $FE_{sm}$ (%) | 2,3 | 2,4 | 4,3 | ↑ |
| SMP (%) | 97,7 | 97,6 | -0,1 | ↓ |

*\* The Central Bank of the Russian Federation (URL: http://www.cbr.ru), \*\* at the end of the year, \*\*\* billion Rubles*

Thus, the securities market is not a source of development of the country's economy, since it does not perform the functions of accumulation and redistribution of funds. This proves that the system is closed and inaccessible to the broad masses of participants, and again emphasizes the need to develop a step-by-step plan for the development of the Russian securities market.

## 5. Conclusion

Despite the fact that Russia has a significant potential for the development of the stock market, the performance of its securities market and its functions is not adequate in comparison with other developed countries. The improvement of this market will not only contribute to the further development of the financial market, but also greatly contribute to economic development. As a part of the financial system, the Russian securities market requires more attention from the government. Another significant obstacle in its development is the problem of the population's perception of the securities market as one of the sources of income, which was also highlighted in the works of Podgorny (2016, 2017). In other words, solving this problem in Russia presupposes a change in the status of the securities market itself - it should become an effective instrument of the state's financial policy, and not just be a platform for speculative transactions. For this, it is necessary to obtain an objective assessment of the efficiency of the market function.

Most of the modern work in this area is based on the analysis of various indicators of the securities market (such as trading volumes, price dynamics, indices, volatility, etc.). In the absence of initiative on the part of governing structures, the functioning of this market meets mainly the needs of its large participants, which means that there is no incentive for its further development. There is a violation of the basic functions of the stock market. This concerns the process of capital redistribution: the bulk of the profit settles with systemically important participants, the rights of other market participants are violated, as a result, the proportions of the development of industries are distorted, etc., which, in turn, has a direct negative impact on the economy.

At the heart of building a strategy for managing the securities market should be a system of indicators, characterizing the efficiency of its functioning. It should be built in such a way that the securities market becomes an effective instrument of financial policy, which can ensure the development of the country's economy as a whole. To this aim, there are several actions, including:

1. Reducing the negative impact of external factors (political and social instability, prevention of economic and economic crises).

2. Redistribution of available free cash resources in the industry, contributing to the restoration and development of production.

3. Increasing the volume of the domestic stock market.

4. Improvement of the legal framework of intelligent supervision.

5. Increasing the role of the government in the stock market.

6. Protecting investors from losses by creating a public-private system.

7. Implementation of the principle of transparency of information by expanding the volume of publications on the activities of securities issuers.

8. Involvement in industrial turnover of an increasing number of shares of joint stock firms.

The methodology proposed in this paper allows us to assess the degree to which emerging securities markets perform their functions on the basis of publicly available information, published by official sources, which is a significant advantage. The objectivity of the result obtained, in our opinion, is due to the use of indicators based on the values of monetary aggregates. The study of generally accepted indicators, which characterize the dynamics of the securities market does not allow assessing the effectiveness of its development, whereas taking into account the values of the money supply makes it possible to not only track changes in its state, but to also assess the impact on the country's economy as a whole. Thus, the presented study proposes a new approach to assess the efficiency of the securities market function.

The limitations of the proposed method include underlying indicators do not take into account possible receipts from foreign participants, and therefore measure only the internal potential of the securities market of a particular country. The purpose of this work is only to the degree to which the securities market fulfils its main functions of accumulating and distributing investors' funds. Outside Russia, it means mainly the population of the country, while in Russia the process of investing in the securities market to overcome its isolation, has yet to be developed. Although, this methodology can be adapted for the study of emerging markets in other countries, and used by both individuals and regulators to determine the effectiveness of policies for the development of financial markets and the country's economy as a whole. Another limitation of this study includes the amount of available information on the dynamics of financial indicators. The direction of future research is the application of the developed methodology to check the effectiveness of the performance of functions by other financial institutions of the government: budgetary and extra-budgetary funds, state and municipal organizations, and other monetary systems.

Challenges and threats. *Economy of Region*, 14, 4.

Khairov, B.G., and Khasanov, R.Kh. (2016). Research of the efficiency of the Russian stock market as a tool for attracting capital. *Bulletin of the Siberian Institute of Business and Information Technologies*, 3 (19), 95-100.

Konovalova, M.E., and Kuzmina, O.Yu. (2019). Stock crises and their early diagnosis. *Economic theory*, 12 (181), 22-25.

Kuncl, M. (2019). Securitization under asymmetric information over the business cycle. *European Economic Review*, 111, 237-256.

Lavrenova, E.S., and Ilyina, T.G. (2020). A note on the predictability of the Russian stock market. *Bulletin of the Tomsk State University. Economics*, 49, 160-182.

Li, R., Wang, X., Yan, Z., and Zhang, Q. (2019). Trading against the grain: When insiders buy high and sell low. *Journal of Portfolio Management*, 46, 1, 139-151.

Mazaev, N.Yu. (2018). Application of in-country macroeconomic methodology to predict the dynamics of the Russian stock market during the recovery from the recession in 2016-2017. *Modern science: topical problems of theory and practice. Series: Economics and Law*, 2, 60-64.

Mityai, S.A. (2019). Emerging financial markets in the IER system. Available at: Literus.narod.ru/Bussines/MirEcon/3-g1-8.htm

Nayak, R.K., Tripathy, R., Mishra, D., Burugari, V.K., Selvaraj, P., Sethy, A., and Jena, B. (2019). Indian stock market prediction based on rough set and support vector machine approach. *Smart Innovation, Systems and Technologies*, 153, 345-355.

Omran, Sh. (2017). Analysis of the efficiency of the Russian stock market. *Bulletin of the Plekhanov Russian University of Economics*, (6), 90-95.

Podgorny, B.B. (2016). The attitude of the population to the stock market - a comparative factual study. *News of the South-West State University*, 4 (21), 209-217.
35